# Negative differential photoconductance in InGaN/GaN multiple-quantum-well structures

M.V. Baranovskiy[*], G.F. Glinskii, and M.S. Mironova

*Department of Micro- and Nanoelectronics, Saint Petersburg Electrotechnical University "LETI",
197376 St. Petersburg, Russian Federation*

[*]E-mail: mv.baranovskiy@gmail.com

**Abstract** The photocurrent and the small-signal photoconductance of InGaN/GaN multiple-quantum-well structures were studied at the temperature range from 10 to 300 K. The optical excitation was carried out at the quantum wells intrinsic absorption wavelengths. Regardless of the temperature the experimental plots of direct photocurrent vs. reverse voltage were step-like, which is related to the sequential quantum wells passage from quasi-neutrality into the *p-n*-junction space charge region. In addition, under optical excitation near the quantum wells material absorption edge we observed the photocurrent declines with increasing reverse bias, i.e. the negative differential photoconductance. This phenomenon is associated with a blue shift of the InGaN quantum well absorption edge arising due to compensation of its build-in piezoelectric field by the *p-n*-junction electric field. Furthermore, it was experimentally shown that each quantum well corresponds to two peaks in the small-signal photoconductance vs. reverse voltage dependence. The temperature changes in the amplitude and position of these peaks indicate that they probably related to the charge carriers thermal emission and thermally activated tunneling from the quantum well.

Heterostructures based on gallium, indium and aluminum nitrides are widely used for the creation of light-emitting diodes[1,2], lasers[3], photodetectors[4], high-frequency transistors[5] and other devices. Such structures are grown on foreign substrates, mainly on the *c*-plane sapphire. Different layers of the structure are usually strongly lattice-mismatched. For these reasons, group-III nitride heterostructures characterized by high dislocation density, strong mechanical stress and piezoelectric field, which make them much different from the ideal heterostructures. Therefore, the actual problem is the experimental study of the properties of group-III nitride heterostructures and in particular quantum well structures.

Currently, the most widespread methods for investigation of quantum well structures are the photo- and electroluminescence spectroscopy[6-8], the photo- and electroreflectance spectroscopy[9-11]. These methods allow determining the band structure features, charge carriers energy spectrum, built-in electric fields and other important parameters. No less informative are the capacitance methods. Capacitance-voltage profiling[12,13] is used to estimate the charge carriers spatial distribution in the heterostructure, which allows determining the quantum well number and the distance between them. The band offsets on heterointerfaces, the energy spectrum and wave functions of charge carriers in quantum wells can also be identified by the simulation of capacitance-voltage characteristics using self-consistent solution of Schrödinger and Poisson equations[14,15]. The rates of emission and capture of charge carriers by deep levels, quantum wells and dots can be obtained by the methods of Deep-level transient spectroscopy[16,17] and the admittance spectroscopy[18,19]. The photocurrent, photovoltage and photocapacitance spectroscopies[20-22] are the most common among photovoltaic methods. As an electrical signal proportional to the number of absorbed photons, the photocurrent spectrum reflects the features of the absorbance. Therefore, the photovoltaic methods, like the optical, allow to analyze the semiconductors band structure features and the embedded fields value. Apart from this, they provide information on the charge carrier transport mechanisms[23].

This paper describes investigations of the photocurrent and differential photoconductance of light-emitting diodes (LEDs) based on InGaN/GaN multiple-quantum-well (MQW) structures. We observed a pronounced step character of the photocurrent on the reverse bias dependence. It is due to the depletion region boundary sequentially passing through the quantum wells, which is confirmed by the results of capacitance-voltage measurements. Furthermore, when excited by light with photon energy close to the quantum wells material absorption edge, we have found the photocurrent declines with increasing reverse bias, i.e. the negative differential photoconductance. This phenomenon we associate with the fundamental absorption edge shift that occurs when the piezoelectric field in the quantum well is partially compensated by the field of *p-n*-junction. For each quantum well we also observed two peaks in the differential photoconductance dependence on reverse voltage, the relative magnitude of which varies significantly with temperature and frequency. We assume that these peaks correspond to different mechanisms of charge carrier transport.

Studies were conducted on the LEDs based on the GaN *p-n*-junction grown on *c*-plane sapphire substrates by metal-organic chemical vapor deposition (MOCVD). The structures contain multiple InGaN quantum wells, concluded between the wide-band GaN barriers. Capacitance-voltage profiling revealed five sharp concentration peaks at a distance of approximately 18 nm from each other (Fig. 1), which correspond to five quantum wells. The *p-n*-junction area is 0.067 mm$^2$. The LEDs emit in the blue range, the wavelength at the maximum is 465 nm.

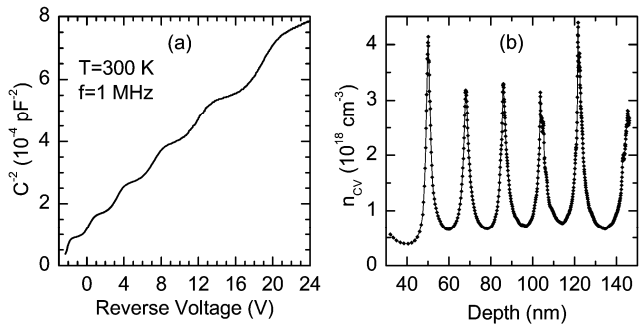

FIG. 1. (a) Experimental dependence of $1/C^2$ vs. reverse voltage for InGaN/GaN MQW structure. (b) Depth profile of the apparent electron concentration.

The measurements were carried out on an automated setup consisting of a closed cycle helium cryostat Janis CCS-150/204N, temperature controller LakeShore 325, immittance meter E7-20, programmable power supply Agilent E3643A and the computer. The measurements of photovoltaic characteristics were performed with constant exciting light flux from blue or violet LEDs. Wavelengths at the emission maximum are 465 nm for the blue LED (same as the samples emission spectra) and 415 nm for the violet LED. The spectral line width is approximately 30 nm in both cases. Such a wide excitation spectrum does not lead to a significant distortion of the measured characteristics which was monitored by measurements with monochromatic excitation at the same wavelengths. However, the high optical power of the LEDs (about 1 mW), in contrast to a monochromatic source, significantly increases the signal to noise ratio, allowing to dispense with lock-in technique.

The structures optical excitation was carried out through the top *p*-layer. Since the photon energy is greater than the InGaN band gap and at the same time does not exceed the GaN band gap, the light is absorbed only in the quantum wells, but not in the barriers. Due to the small width of the quantum wells a fraction of the absorbed light is small, so we can assume that the intensity changes insignificantly from well to well, and all quantum wells are excited under the same conditions.

The measurements of the photocurrent and the small-signal photoconductance were carried out. In the first case, a constant reverse bias was applied to the sample and the direct current through the structure was measured. Photocurrent was calculated as the difference between the currents under illumination and in the dark. For the studied samples the dark current did not exceed the units nA and was negligible compared to the photocurrent.

To measure the differential photoconductance a constant reverse bias and a small measuring signal at a predetermined frequency were applied to the sample. The active and the reactive current components were measured to determine the differential conductance and capacitance. The photoconductance was calculated as the difference between the conductances under illumination and in the dark. In the studied samples the dark conductance is negligible at low frequencies, becomes comparable to the photoconductance at frequencies of the order of several kHz, and plays a decisive role in the high frequency region. Our studies were carried out in the range from 100 Hz to 10 kHz, where the effect of photoexcitation on conductance is the most significant.

The experimental plots of the photocurrent vs. the reverse bias for InGaN/GaN multiple-quantum-well structures are shown in Fig. 2. In the case of excitation with violet light (Fig. 2a), the dependence of the photocurrent on voltage has a pronounced step character. Similar features have previously been observed by T.-S. Kim et al.[13], J. Allam et al.[24], C. Rivera et al.[25] and H. M. Khalil et al.[26,27] in quantum-well structures based on various materials (InGaN/GaN, GaInAs/InP, GaInNAs/GaAs). Steps in the photocurrent associated with the sequential passage of individual quantum wells from the quasi-neutrality area into the space charge region. Photocurrent changes can be explained by the features of charge carrier transport as well as a change in absorption. In Ref. 25 it was noted that in such structures absorbance at close wavelengths is changed inessential with voltage. Therefore, the observed phenomenon is caused by transport features of excess charge carriers in the quantum wells.

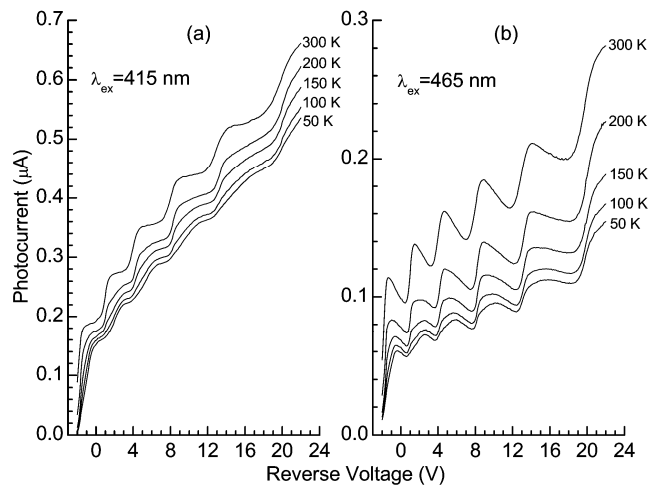

FIG. 2. Photocurrent vs. reverse voltage measured at different temperatures. Optical excitation wavelengths $\lambda_{ex}$ are (a) 415 nm and (b) 465 nm.

While a given quantum well is not affected by the electric field of *p-n*-junction, the main part of light generated charge carriers recombine within the quantum well. A small part of them can overcome the potential barrier and leave the quantum well, but in the steady state the excess electrons and holes move in opposite directions with the same probability, and eventually recombine. Hence, such quantum well does not contribute to the photocurrent.

When the quantum well is inside the space charge region, the electric field of *p-n*-junction breaks the symmetry: electrons move predominantly in the direction of the *n*-region, and holes towards the *p*-region. Thus, such quantum well becomes a source of the photocurrent, which value is determined by the number of charge carriers, extracted from the quantum well by an electric field. At sufficiently high fields almost all charge carriers generated by the light per unit of time are pulled out of the quantum well, and the photocurrent is saturated.

When excited by blue light (Fig. 2b), the dependence of photocurrent on voltage is significantly different from the previous case. We observed the photocurrent declines with reverse bias increasing, i.e. the negative differential photoconductance. An essential condition for the occurrence of this phenomenon is that the optical excitation is carried out near the absorption edge of the quantum wells material InGaN. As discussed in our previous work[28], this behavior is associated with a decrease in the light absorption in a given quantum well as it moves deeper into the depletion region of the *p-n*-junction. This phenomenon is illustrated in Fig. 3. At small values of the reverse bias (Fig. 3a), when the InGaN quantum well is in the quasi-neutrality region, it has a strong built-in piezoelectric field. Due to quantum-confined Stark effect, the energy gap between the ground states of electrons and holes in the quantum well is significantly less than in a similar quantum well without the piezoelectric field. With increasing reverse bias (Fig. 3b) quantum well passes into the *p-n*-junction space charge region, the electric field of which partially compensates the quantum well piezoelectric field. The ground-state transition energy is increased, and hence the optical absorption edge shifts to shorter wavelengths. Obviously, this effect is significant only when excited near the absorption edge. As a result, the number of absorbed photons per unit time is considerably reduced at the wavelength of 465 nm, while remains practically unchanged at the wavelength of 415 nm.

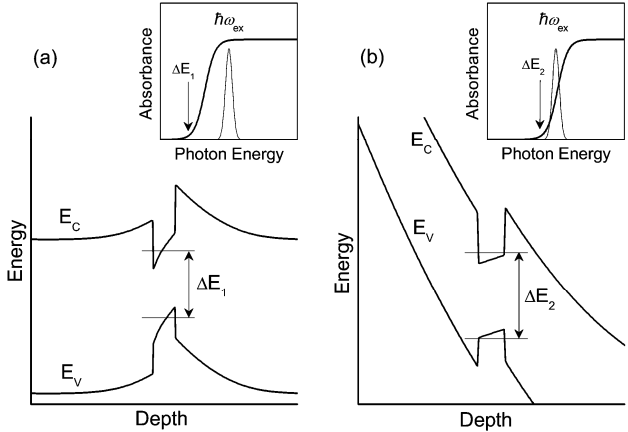

FIG. 3. Schematic energy diagram of a quantum well (a) in the quasi-neutrality region and (b) in the *p-n*-junction space-charge region. The insets show the position of the excitation line relative to the absorption edge.

It can be noted that the sharpest increase in the photocurrent with reverse voltage is observed at the room temperature, and it becomes smoother at lower temperatures. This is because the charge carriers are emitted from the quantum wells by the thermal emission and the thermally activated tunnelling. Obviously, electrons and holes photogenerated in the quantum well must overcome the potential barrier and reach the wide-band region of the structure to contribute the photocurrent. The greater the ratio of the emission rate to the recombination rate, the higher the portion of the ejected charge carriers from the total number of carriers, generated by light per unit time. With decreasing temperature the emission rate is falling and the part of the recombining carriers increases. To restore the previous balance between ejected and recombining carriers the higher electric field required. Therefore, as the temperature decreases the photocurrent saturation areas (Fig. 2a) are shifted to higher voltages.

Thus, the measurements of the photocurrent vs. applied reverse voltage can be used for rapid determination of the amount of quantum wells in the structure, the distance between them, the quality of heterointerfaces and other parameters[29,30]. But more detailed information can be obtained from the small-signal photoconductance studies[31]. At the frequency tending to zero this quantity makes sense of the photocurrent derivative with respect to the voltage. However, at non-zero frequency additional features associated with the finite time of the space charge variation may appear. Fig. 4 shows the plots of differential photoconductance measured at different temperatures and frequencies vs. reverse bias. At high temperatures (the upper graphs) there is a series of sharp peaks. Each of them corresponds to an increase in photocurrent at the voltage where one of the quantum wells enters the space charge region. With decreasing temperature (downwards on Fig. 4), the amplitude of these narrow peaks falls. However, an additional series of broad peaks appears, the amplitude of which, on the contrary, grows with decreasing temperature.

We assume that these features are associated with different mechanisms of carriers escape from the quantum wells into the wide-band barriers. Narrow peaks correspond to the thermal emission. A number of charge carriers per unit of time escaping quantum well by this way does not depend on the applied voltage, and is determined only by the potential barrier height and temperature. But they begin to contribute to the photocurrent only when the quantum well passes into the depletion region of *p-n*-junction. This causes the narrowness of the observed peaks.

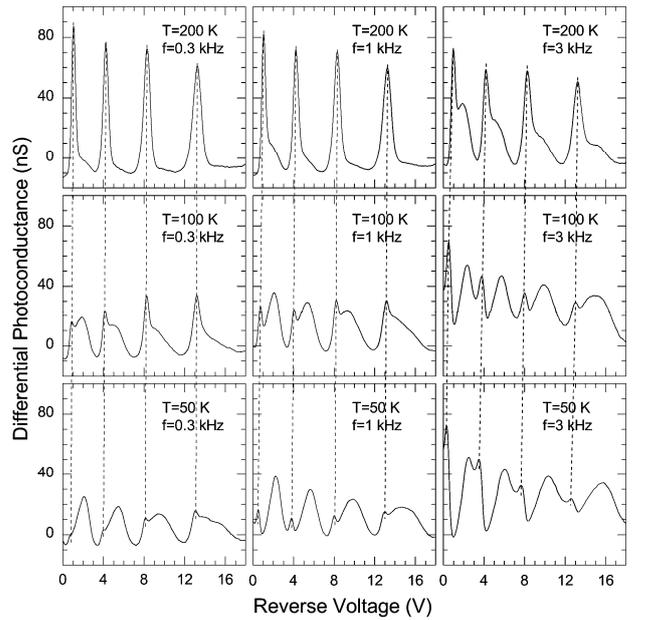

FIG. 4. Differential photoconductance vs. reverse voltage at different temperatures (top to bottom) and frequencies (left to right). Optical excitation wavelength $\lambda_{ex}$=465 nm.

At low temperatures the thermally activated tunneling prevails. This mechanism involves the charge carriers thermal transition to the excited energy states and their subsequent tunneling through the potential barrier. The higher the charge carrier energy state and the greater the electric field, the narrower potential barrier and, thus, higher probability of tunneling. Therefore, the emission rate in this case is determined by the temperature and the electric field in the quantum well. Increase of the electric field leads to the emission rate gradual growth and, hence, the photocurrent growth, so the second series peaks are sufficiently broad. As mentioned above, when the temperature decreases charge carriers ejection from quantum wells occurs at higher voltages, which explains the broad peaks temperature shift.

Note that the amplitude and position of the conductance peaks depends on the frequency. Increased frequency at a constant temperature (left to right on Fig. 4) leads to a marked increase in the amplitude of the broad peaks and their shift toward higher voltages, while narrow peaks, in contrast, are slightly shifted toward smaller voltages. Probably a significant role in this phenomenon plays a ratio of the frequency to the emission rates of charge carriers from the quantum wells.

In conclusion, we investigated the photocurrent and differential photoconductance in InGaN/GaN multiple-quantum-well structures. The dependence of the photocurrent on the reverse bias reflects the structure of the sample under study, so such measurements may be used to determine the amount of quantum wells, the distance between them, the quality of heterointerfaces and other parameters. It was experimentally shown that under optical excitation near the quantum wells material absorption edge the negative differential photoconductance appears. This phenomenon is associated with the absorption edge blue shift arising due to compensation of quantum wells build-in piezoelectric field by the electric field of the *p-n*-junction. It was found that each quantum well corresponds to two peaks in the small-signal photoconductance. Apparently, they are associated with different mechanisms of the charge carriers escape from the quantum well. The temperature changes in the amplitude and position of these peaks indicate that these mechanisms are the thermal emission and the thermally activated tunneling.